\documentclass[sigconf,nonacm]{acmart}

\usepackage{subfigure}
\usepackage{siunitx}
\usepackage{enumitem}

\usepackage{algorithm}
\usepackage[noend]{algpseudocode}
\let\ReturnInline\Return
\renewcommand{\Return}{\State\ReturnInline}
\algrenewcommand\algorithmicrequire{$\rhd$}
\algrenewcommand\algorithmicensure{$\square$}

\AtBeginDocument{%
  \providecommand\BibTeX{{%
    \normalfont B\kern-0.5em{\scshape i\kern-0.25em b}\kern-0.8em\TeX}}}

\setcopyright{acmcopyright}
\copyrightyear{2018}
\acmYear{2018}
\acmDOI{XXXXXXX.XXXXXXX}

\acmConference[Conference acronym 'XX]{Make sure to enter the correct
  conference title from your rights confirmation emai}{June 03--05,
  2018}{Woodstock, NY}

\newcommand{\ignore}[1]{}

\newcommand{\ours}{$\nu$-LPA}

\begin{document}

\title[\ours{}: Fast GPU-based Label Propagation Algorithm (LPA) for Community Detection]{\ours{}: Fast GPU-based Label Propagation Algorithm (LPA) \\for Community Detection}


\author{Subhajit Sahu}
\email{subhajit.sahu@research.iiit.ac.in}
\affiliation{%
  \institution{IIIT Hyderabad}
  \streetaddress{Professor CR Rao Rd, Gachibowli}
  \city{Hyderabad}
  \state{Telangana}
  \country{India}
  \postcode{500032}
}


\settopmatter{printfolios=true}

\begin{abstract}
Community detection is the problem of identifying natural divisions in networks. Efficient parallel algorithms for identifying such divisions are critical in a number of applications. This report presents an optimized implementation of the Label Propagation Algorithm (LPA) for community detection, featuring an asynchronous LPA with a Pick-Less (PL) method every $4$ iterations to handle community swaps, ideal for SIMT hardware like GPUs. It also introduces a novel per-vertex hashtable with hybrid quadratic-double probing for collision resolution. On an NVIDIA A100 GPU, our implementation, \ours{}, outperforms FLPA (sequential), NetworKit LPA (multicore), Gunrock LPA (GPU), and cuGraph Louvain (GPU) by $364\times$, $62\times$, $2.6\times$, and $37\times$, respectively, while running FLPA and NetworKit LPA on a server with dual 16-core Intel Xeon Gold 6226R processors --- processing $3.0B$ edges/s on a $2.2B$ edge graph --- and achieves $4.7\%$ higher modularity than FLPA, but $6.1\%$ and $9.6\%$ lower than NetworKit LPA and cuGraph Louvain.
\end{abstract}



\maketitle

\section{Introduction}
\label{sec:introduction}
Community detection involves identifying groups of vertices within a network that have stronger connections to each other than to the rest of the network \cite{com-fortunato10}. When these communities are determined solely from the network's structure, they are called intrinsic, and if each vertex belongs to only one group, the communities are considered disjoint \cite{com-gregory10, coscia2011classification}. These groups, often referred to as clusters, provide valuable insights into the network's structure and behavior \cite{com-fortunato10, abbe2018community}. Applications of community detection span diverse fields, including ecology \cite{guimera2010origin}, healthcare \cite{salathe2010dynamics, bechtel2005lung, haq2016community}, machine learning \cite{bai2024leiden, das2011unsupervised}, drug discovery \cite{ma2019comparative, udrescu2020uncovering}, analysis of biological networks \cite{kim2009centralized, rivera2010nemo, popa2011directed}, exploration of human brain networks \cite{he2010graph}, and various other graph-related problems \cite{stergiou2018shortcutting, meyerhenke2017parallel, slota2020scalable, henne2015label, gottesburen2021scalable, boldi2011layered, valejo2020coarsening}.

Community detection is challenging because there is no prior knowledge about the number of communities or their size distribution \cite{com-gregory10}. As a result, heuristic-based methods are often used to address this task \cite{com-raghavan07, com-blondel08, com-xie11, traag2023large, clauset2004finding, duch2005community, reichardt2006statistical, com-kloster14, com-traag19, com-you20, com-rosvall08, com-whang13}. The quality of detected communities is commonly evaluated using the modularity metric \cite{com-newman06}. Among these methods, the Label Propagation Algorithm (LPA), also known as RAK \cite{com-raghavan07}, is a popular diffusion-based heuristic that efficiently detects communities of moderate quality. LPA stands out for its simplicity, speed, and scalability compared to the Louvain method \cite{com-blondel08}, another prominent algorithm recognized for producing high-quality communities. Here, we observe that LPA is $37\times$ faster than Louvain, although the communities it identifies have a quality that is, on average, $9.6\%$ lower. Although LPA tends to produce communities with lower modularity scores, it has been shown to achieve high Normalized Mutual Information (NMI) relative to ground truth \cite{peng2014accelerating}. In our evaluation of other label-propagation-based methods such as COPRA \cite{com-gregory10}, SLPA \cite{com-xie11}, and LabelRank \cite{com-xie13}, LPA emerged as the most efficient, delivering communities of comparable quality \cite{sahu2023selecting}.

In this report, we present our parallel GPU implementation of LPA\footnote{\url{https://github.com/puzzlef/rak-communities-cuda}}, which we call \ours{}. This implementation builds on our efficient multicore version of LPA, GVE-LPA \cite{sahu2023gvelpa}. In addition to adapting the code for GPU use, we introduced several key optimizations: \textbf{(1)} We employ an efficient hashtable stored in global memory that uses open addressing with a hybrid approach combining quadratic probing and double hashing for collision resolution. \textbf{(2)} To mitigate community swaps --- which can hinder algorithm convergence --- we restrict each vertex to changing its label only if the new label has a smaller ID value than its current label, once every $4$ iterations. \textbf{(3)} We use two distinct kernels to handle low-degree and high-degree vertices: a thread-per-vertex kernel and a block-per-vertex kernel, respectively, and determine an optimal switch degree between the two. \textbf{(4)} Finally, unlike GVE-LPA, we use 32-bit floating-point numbers for hashtable values instead of 64-bit, which improves performance and reduces memory usage without affecting the quality of the detected communities. Apart from Gunrock LPA \cite{wang2016gunrock}, no other functional GPU-based LPA implementations are publicly available. Thus, we compare our performance with state-of-the-art sequential implementations, such as FLPA \cite{traag2023large}, and multicore implementations like NetworKit LPA \cite{staudt2016networkit}. Additionally, we compare \ours{} with cuGraph Louvain \cite{kang2023cugraph} to highlight the trade-offs between the Louvain algorithm and LPA.

\section{Related work}
\label{sec:related}
Label Propagation Algorithm (LPA) has been widely applied in various fields. Das and Petrov \cite{das2011unsupervised} use LPA for cross-lingual knowledge transfer, employing projected labels in an unsupervised part-of-speech tagger for languages lacking labeled training data. Wang et al. \cite{wang2013label} extend LPA to propagate labels from 2D semantic labeled datasets like ImageNet to 3D point clouds, addressing challenges in acquiring sufficient 3D labels for training classifiers. LPA has also proven effective in automatic segmentation tasks \cite{wang2014geodesic}. Aziz et al. \cite{aziz2023novel} apply a modified LPA for sectionalizing power systems, while Stergiou et al. \cite{stergiou2018shortcutting} propose Shortcutting Label Propagation for distributed connected components. Boldi et al. \cite{boldi2011layered} introduce Layered Label Propagation, which uses node clusterings in various layers to reorder and compress graph nodes with the WebGraph compression framework \cite{boldi2004webgraph}. Valejo et al. \cite{valejo2020coarsening} present a weight-constrained variant of LPA for fast graph coarsening, allowing users to specify network size and control super-vertex weights. Mohan et al. \cite{mohan2017scalable} propose a scalable method for community structure-based link prediction on large networks, combining parallel LPA for community detection with a parallel Adamic–Adar measure for link prediction. Xu et al. \cite{xu2019distributed} develop a distributed temporal link prediction algorithm, DTLPLP, which updates node labels based on the weights of incident links and aggregates values from similar source nodes for link score evaluation. LPA has also been widely applied in graph partitioning \cite{meyerhenke2014partitioning, meyerhenke2016partitioning, zhang2020multilevel, slota2020scalable, wang2014partition, akhremtsev2020high, slota2014pulp, meyerhenke2017parallel, bae2020label}.

Significant work has been conducted to improve the original LPA through various modifications. Farnadi et al.'s \cite{farnadi2015scalable} Adaptive Label Propagation dynamically adjusts to network characteristics, like homophily or heterophily. Shen and Yang's \cite{shen2016topic} simLPA combines content-based and link-structure methods. Zarei et al.'s \cite{zarei2020detecting} Weighted Label Propagation Algorithm (WLPA) for signed and unsigned networks utilizes MinHash to estimate node similarity. Sattari and Zamanifar \cite{sattari2018spreading} addressed "monster communities" with a label-activation strategy. Zheng et al. \cite{zheng2018improved} introduced label purity, prioritized nodes with higher weighted degrees, and used an attenuation factor for faster convergence. Zhang et al.'s \cite{zhang2017label} LPA\_NI ranks nodes by importance, for label updates, while Berahmand and Bouyer's \cite{berahmand2018lp} Label Influence Policy for Label Propagation Algorithm (LP-LPA) evaluates connection strengths and label influence --- in order to enhance stability. El Kouni et al.'s \cite{el2021wlni} WLNI-LPA integrates node importance, attributes, and network topology to improve labeling accuracy. Roghani et al.'s \cite{roghani2021pldls} Parallel Label Diffusion and Label Selection (PLDLS) identifies core nodes through triangle formation, and proceeds with standard LPA iterations. Li et al.'s \cite{li2015parallel} Parallel Multi-Label Propagation Algorithm (PMLPA) uses an ankle-value based label updating strategy to detect overlapping communities. Finally, Zhang et al.'s \cite{zhang2023large} recent method employs core nodes with degrees above the average, propagating labels layer-by-layer for detecting overlapping communities.

Some GPU-based implementations of LPA have also been proposed. Soman and Narang \cite{soman2011fast} presented a parallel GPU algorithm for weighted label propagation. Kozawa et al. \cite{kozawa2017gpu} proposed a GPU-accelerated implementation of LPA, and discussed algorithms for handling datasets which do not fit in GPU memory. Ye et al. \cite{ye2023large} recently developed GLP, a GPU-based framework for label propagation. However, these implementations are either unavailable due to restrictions (e.g., company policies) \cite{soman2011fast, ye2023large} or fail to run due to runtime issues, such as errors when loading large graphs \cite{kozawa2017gpu}.

A few open-source implementations for community detection using LPA have been developed. Fast Label Propagation Algorithm (FLPA) \cite{traag2023large} is a fast variant of LPA that uses a queue-based approach to process only vertices with recently updated neighborhoods, without random node order shuffling. NetworKit \cite{staudt2016networkit} is a software package for analyzing large graph datasets, implemented with C++ kernels and a Python frontend, and includes a parallel LPA implementation. It employs a boolean flag vector for active node tracking, uses OpenMP's guided schedule for parallel processing, and uses \texttt{std::map} for storing label weights. We recently introduced GVE-LPA \cite{sahu2023gvelpa}, an efficient multicore implementation of LPA that outperforms FLPA/NetworKit LPA by $139\times$/$40\times$.

Despite the utility of LPA, there is a lack of efficient and widely available GPU-based implementations, to the best of our knowledge. Our proposed \ours{} attempts to fill this gap.

\section{Preliminaries}
\label{sec:preliminaries}
Consider an undirected graph $G(V, E, w)$, where $V$ is the set of vertices, $E$ is the set of edges, and $w_{ij} = w_{ji}$ is the weight for each edge. For an unweighted graph, each edge has a unit weight, $w_{ij} = 1$. The neighbors of a vertex $i$ are denoted by $J_i = \{j \ | \ (i, j) \in E\}$, and the weighted degree of vertex $i$ is $K_i = \sum_{j \in J_i} w_{ij}$. The graph has $N = |V|$ vertices, $M = |E|$ edges, and the total sum of edge weights in the undirected graph is $m = \sum_{i, j \in V} w_{ij} / 2$.

\subsection{Community detection}
\label{sec:about-communities}

Disjoint community detection aims to map each vertex $i \in V$ to a community-id $c$ from a set $\Gamma$, using a community membership function $C: V \rightarrow \Gamma$. The set of vertices in community $c$ is denoted as $V_c$, and the community to which vertex $i$ belongs is $C_i$. For a vertex $i$, its neighbors in community $c$ are represented as $J_{i \rightarrow c} = \{j\ |\ j \in J_i \text{ and } C_j = c\}$, and the sum of the corresponding edge weights is $K_{i \rightarrow c} = \sum_{j \in J_{i \rightarrow c}} w_{ij}$. The total weight of edges within a community $c$ is $\sigma_c = \sum_{(i, j) \in E \text{ and } C_i = C_j = c} w_{ij}$, while the total edge weight associated with $c$ is $\Sigma_c = \sum_{(i, j) \in E \text{ and } C_i = c} w_{ij}$\ignore{\cite{com-leskovec21}}.

\subsection{Modularity}
\label{sec:about-modularity}

Modularity is a fitness metric for assessing the quality of communities found by heuristic-based community detection algorithms, calculated as the difference between the fraction of edges within communities and the expected fraction if edges were randomly distributed, ranging from $[-0.5, 1]$, with higher values indicating better outcomes \cite{com-brandes07}. The modularity $Q$ of detected communities is determined using Equation \ref{eq:modularity}, involving the Kronecker delta function ($\delta(x,y) = 1$ if $x=y$, $0$ otherwise). Additionally, the \textit{delta modularity} for transferring a vertex $i$ from community $d$ to $c$, represented as $\Delta Q_{i: d \rightarrow c}$, is computed using Equation \ref{eq:delta-modularity}.

\begin{equation}
\label{eq:modularity}
  Q
  = \frac{1}{2m} \sum_{(i, j) \in E} \left[w_{ij} - \frac{K_i K_j}{2m}\right] \delta(C_i, C_j)
  = \sum_{c \in \Gamma} \left[\frac{\sigma_c}{2m} - \left(\frac{\Sigma_c}{2m}\right)^2\right]
\end{equation}

\begin{align}
\begin{split}
\label{eq:delta-modularity}
  &\Delta Q_{i: d \rightarrow c} = \Delta Q_{i: d \rightarrow i} + \Delta Q_{i: i \rightarrow c} \\
  &= \left[ \frac{\sigma_d - 2K_{i \rightarrow d}}{2m} - \left(\frac{\Sigma_d - K_i}{2m}\right)^2 \right] + \left[ 0 - \left(\frac{K_i}{2m}\right)^2 \right] - \left[ \frac{\sigma_d}{2m} - \left(\frac{\Sigma_d}{2m}\right)^2 \right] \\
  &+ \left[ \frac{\sigma_c + 2K_{i \rightarrow c}}{2m} - \left(\frac{\Sigma_c + K_i}{2m}\right)^2 \right] - \left[ \frac{\sigma_c}{2m} - \left(\frac{\Sigma_c}{2m}\right)^2 \right] - \left[ 0 - \left(\frac{K_i}{2m}\right)^2 \right] \\
  &= \frac{1}{m} (K_{i \rightarrow c} - K_{i \rightarrow d}) - \frac{K_i}{2m^2} (K_i + \Sigma_c - \Sigma_d)
\end{split}
\end{align}

\subsection{Label Propagation Algorithm (LPA)}
\label{sec:about-rak}

LPA \cite{com-raghavan07} is a widely used diffusion-based technique for detecting communities of moderate quality in large-scale networks. It is simpler, faster, and more scalable compared to the Louvain method \cite{com-blondel08}. In LPA, each vertex $i$ initially has a unique label (community ID) $C_i$. During each iteration, vertices update their labels by adopting the label with the highest total interconnecting weight, as described in Equation \ref{eq:rak}. This iterative process continues until densely connected groups of vertices reach a consensus, effectively forming communities. The algorithm terminates when at least a fraction of $1-\tau$ of the vertices (with $\tau$ being the tolerance parameter) retain their current labels. LPA has a time complexity of $O(L |E|)$ and a space complexity of $O(|V| + |E|)$, where $L$ represents the number of iterations performed by the algorithm \cite{com-raghavan07}.

\begin{equation}
\label{eq:rak}
  C_i =\ \underset{c\ \in \ \Gamma}{\arg\max} { \sum_{j \in J_i\ |\ C_j = c} w_{ij} }
\end{equation}

\subsection{Open Addressing in Hashing}

Open addressing, also known as closed hashing, is a technique for resolving collisions in hash tables. Instead of using additional data structures like linked lists (as in separate chaining), open addressing manages all entries within a single array. When a hash collision occurs, it is resolved by probing, or searching through alternative positions in the array --- the probe sequence --- until either the desired record is found or an empty slot is located, indicating that the key is not present in the table \cite{tenenbaum1990data}. Common probe sequences include: \textbf{(1)} \textit{Linear probing}, where the step between probes is fixed (often set to $1$); \textbf{(2)} \textit{Quadratic probing}, where the interval between probes increases according to a quadratic function; and \textbf{(3)} \textit{Double hashing}, which uses a second hash function to determine a fixed interval for each record. These methods present trade-offs: linear probing offers the best cache performance but is highly susceptible to clustering, double hashing exhibits virtually no clustering but has weaker cache performance, and quadratic probing strikes a balance between the two. A key factor affecting the efficiency of an open addressing hash table is the load factor, which is the ratio of occupied slots to the total capacity of the array. As the load factor approaches $100\%$, the number of probes needed to find or insert a key grows significantly. For most open addressing methods, a typical load factor is around $50\%$.

\subsection{Fundamentals of a GPU}

The fundamental building block of NVIDIA GPUs is the Streaming Multiprocessor (SM). Each SM houses multiple CUDA cores for executing parallel threads. Additionally, SMs have shared memory, registers, and specialized function units. The number of SMs varies by GPU model, with each SM operating independently\ignore{to manage multiple parallel threads}. The memory hierarchy of NVIDIA GPUs includes global, shared, and local memory. Global memory is the largest but slowest, shared memory is a low-latency memory shared among threads within an SM, while local memory serves as private storage for individual threads when register space is insufficient \cite{cuda-sanders10, gpu-nickolls10}.

Threads on a GPU are organized differently from those on a multicore CPU, structured into warps, thread blocks, and grids. A warp consists of 32 threads that execute instructions in lockstep. Thread blocks are groups of threads that run on the same SM. Within a thread block, warps execute together, and the SM schedules alternate warps if threads stall, such as when waiting for memory. Threads in a block can communicate using shared memory, a private, user-managed cache within each SM. A grid, which comprises multiple thread blocks, provides a higher-level structure for managing parallelism and optimizing resource usage. Thread blocks within a grid communicate exclusively through global memory, which, while slower than shared memory, allows for data exchange across different blocks \cite{cuda-sanders10, gpu-nickolls10}.

\section{Approach}
\label{sec:approach}
Our GPU-based implementation of LPA builds upon GVE-LPA \cite{sahu2023gvelpa}, incorporating several key features: \textbf{(1)} An asynchronous parallel version of LPA using a single community membership vector, promoting faster convergence but potentially introducing variability in results; \textbf{(2)} A maximum of $20$ iterations; \textbf{(3)} A per-iteration tolerance of $\tau = 0.05$; \textbf{(4)} Vertex pruning to minimize unnecessary computations, where a vertex assigns its neighbors for processing upon label change and is labeled ineligible for further processing once completed; and \textbf{(5)} A strict version of LPA, where each vertex selects the first label with the highest associated weight. We now discuss our GPU-specific techniques for LPA, below.

\subsection{Mitigating community swaps}

We first observe that the GPU implementation of LPA fails to converge for a number of input graphs, and instead continues to run for $20$ iterations. This suggests that several vertices are caught in cycles of community or label swaps. A common scenario that can lead to this issue is when two interconnected vertices continuously adopt each other's labels. In general, this can occur when there is symmetry, i.e., when two vertices are equally connected to each other's community. Community swaps, such as these, are more likely than may be anticipated, due to the fact the GPUs execute in lockstep. Symmetric vertices, such as the ones discussed above, can easily end up swapping labels if assigned to the same SM. Since SM assignment is typically based on vertex IDs, this problem can persist indefinitely, leading to the algorithm’s failure to converge. Hence, introducing symmetry-breaking techniques is crucial.

To address this challenge, we explore two different methods for symmetry breaking. In the \textbf{Cross-Check (CC)} method, after each iteration of LPA, we validate each vertex's community change to ensure that it is a ``good" change. If not, the vertex reverts to its previous community assignment. We consider a community change to be \textit{good} if the new community $c^*$ contains a vertex $i$ with that same ID, i.e., $i = c^*$. In other words, a community is considered to be good if all vertices in the community have joined a certain \textit{leader} vertex, and the leader vertex belongs to the same community. Note how this condition fails in the case of a community swap. Once a ``bad" community change is identified, the community membership of the vertex it atomically reverted to its previous community assignment --- this prevents both vertices involved in the community swap from reverting. In order to be able to perform a revert, we keep of copy of the previous label of each vertex before performing an LPA iteration. We now discuss our \textbf{Pick-Less (PL)} method for tackling the issue of community swaps. In this method, a vertex is only allowed to switch to a new community if the new community ID is lower than its current one. This prevents one of the pairs of vertices involved in a community swap from changing their community membership, thus breaking the symmetry. However, performing either of the two methods too frequently can hinder the algorithm's ability to identify high-quality communities. We experiment with applying either the CC or the PL method every $1$, $2$, $3$, or $4$ iterations of LPA. We do this on large graphs from Table \ref{tab:dataset}, and measure the runtimes, and modularity of obtained communities. We also explore a \textbf{Hybrid (H)} method, combining both the CC and PL methods in all the $16$ possible combinations\ignore{(varying the number of iterations of LPA, after which the CC or PL method is applied)}.

\begin{figure}[hbtp]
  \centering
  \subfigure{
    \label{fig:optswapprevent--runtime}
    \includegraphics[width=0.98\linewidth]{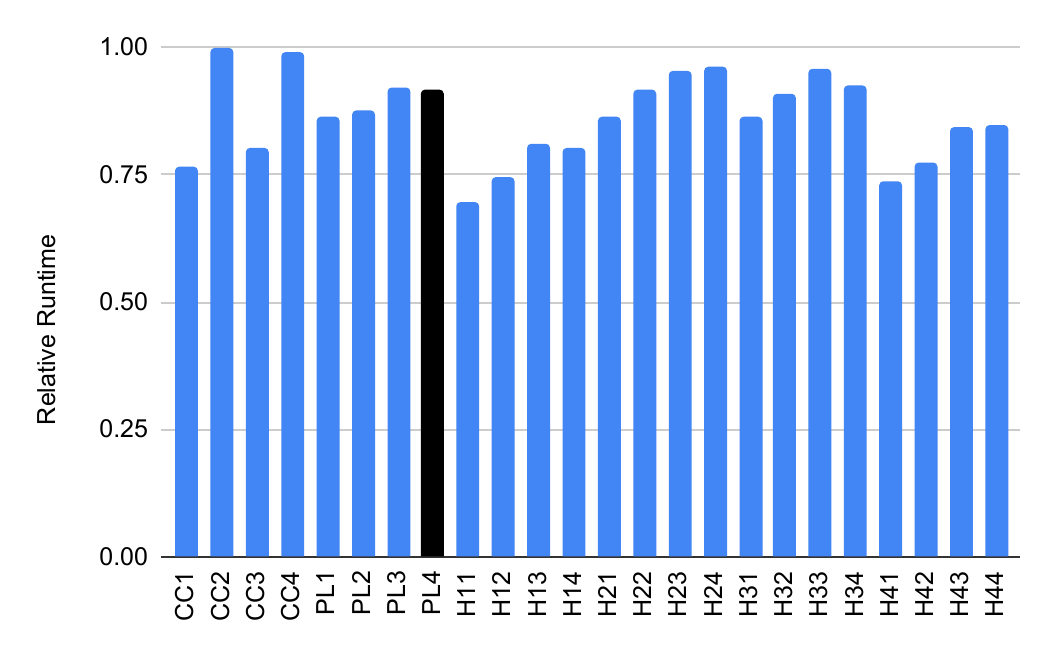}
  }
  \subfigure{
    \label{fig:optswapprevent--modularity}
    \includegraphics[width=0.98\linewidth]{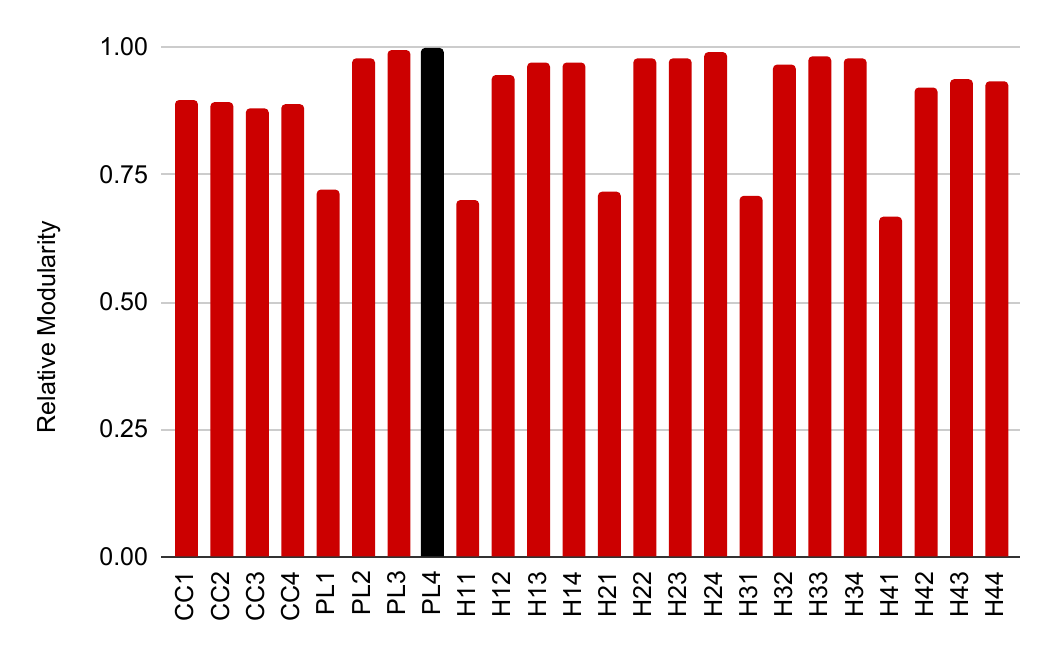}
  } \\[-2ex]
  \caption{Relative Runtime and Modularity of obtained communities with different community swap prevention techniques --- these include cross-checking and reverting bad community swaps (\textit{CC}) every $1$ to $4$ iterations, enforcing picking/selection of only a label with a lower ID value (\textit{PL}) every $1$ to $4$ iterations, and a hybrid of the two techniques (\textit{H}) performed every $1$ to $4$ iterations.}
  \label{fig:optswapprevent}
\end{figure}

Figures \ref{fig:optswapprevent--runtime} and \ref{fig:optswapprevent--modularity} show the mean relative runtime and relative modularity of obtained communities, respectively, for each of the above discussed methods. As the figures show, LPA employing the \textbf{Pick-Less (PL)} method performed every \textbf{4 iterations}, i.e., \textbf{PL4}, identifies communities of the highest modularity, while being only $8\%$ slower that the fastest method (CC2). This leads us to adopt the PL4 method for our GPU implementation of LPA. Note that we employed a double-hashing based hashtable for the above experiment. A discussion on the design of hashtable is given below.

\subsection{Hashtable design}

In GVE-LPA, we utilized per-thread collision-free hashtables, comprising a key list and a full-size values array (of size $|V|$) that were kept well-separated in memory. This design achieved a $15.8\times$ performance improvement compared to C++'s built-in \texttt{std::unordered\_} \texttt{map}. However, unlike multicore CPUs, GPUs operate with a far greater number of threads, making it impractical to allocate $O(|V|)$ memory for each thread on a GPU. This challenge is compounded by the fact that GPUs have limited memory capacity. 

\begin{figure}[hbtp]
  \centering
  \subfigure{
    \label{fig:about-hashtable--all}
    \includegraphics[width=0.86\linewidth]{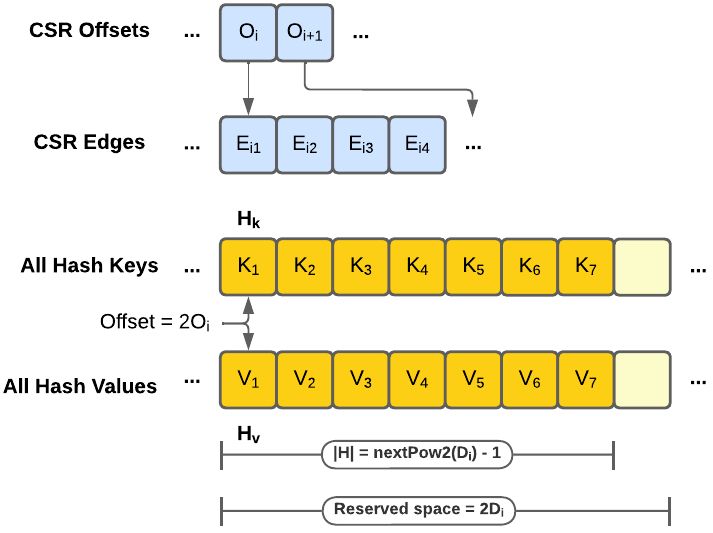}
  } \\[-2ex]
  \caption{Illustration of per-vertex open-addressing hashtables for our GPU implementation of LPA. Each vertex $i$ has a hashtable $H$ with a keys array $H_k$ and a values array $H_v$. Memory for all vertices' hash key and value arrays is allocated together. The hashtable's offset for vertex $i$ is $2O_i$, where $O_i$ is its CSR offset. The memory reserved for the hashtable is $2D_i$, with $D_i$ being the vertex's degree. The hashtable's capacity, or maximum key-value pairs, is $nextPow2(D_i) - 1$.}
  \label{fig:about-hashtable}
\end{figure}

To address these challenges, we instead opt for a hashtable based on open-addressing. This hashtable $H$ is composed of two arrays: a \textit{keys array} $H_k$ and a \textit{values array} $H_v$ --- $H_k$ stores neighboring labels of the current vertex, while $H_v$ holds the corresponding edge weights. The size of the hashtable must be at least as large as the degree of the vertex being processed. However, since a vertex can have a degree as high as $|V|$, allocating a fixed memory block for the hashtable per thread would be impractical, and put us back at square one. Instead, we create a dedicated hashtable for each vertex in the graph. Given that the total number of edges in the graph is $|E|$, the cumulative size of all hashtables can be bounded by $O(|E|)$. In order to ensure a load factor below $100\%$, we allocate twice the degree of each vertex as the size of its corresponding hashtable. This means that memory allocation for all \textbf{per-vertex hashtables} only requires two calls of size $2|E|$: one for all $H_k$s, and another for the $H_v$s. Each thread can then retrieve the hashtable $H$ associated with a vertex, $i$, by computing offsets. This is done using the same offset information provided in the Compressed Sparse Row (CSR) format: specifically, the starting position of the vertex’s edges and its degree. For efficient hash computation, we set the size of each hashtable to $nextPow2(D_i) - 1$, where $D_i$ is the vertex's degree. This enables the use of the $\bmod$ operator as a simple hash function to find the slot index for a given key. Figure \ref{fig:about-hashtable} provides a visual representation of this per-vertex hashtable design. The hashtable supports a few key operations: accumulating weights for matching keys, clearing the hashtable, and identifying the key with the highest associated weight. Deletion of keys is not required for our scenario. To ensure thread safety when accessing and updating the hashtable, we utilize atomic operations. For example, when accumulating weights, we use atomic addition to update $H_k$ and $H_v$. This prevents race conditions when multiple threads attempt to modify the same key's weight simultaneously.

\begin{figure}[hbtp]
  \centering
  \subfigure{
    \label{fig:optresolution--runtime}
    \includegraphics[width=0.98\linewidth]{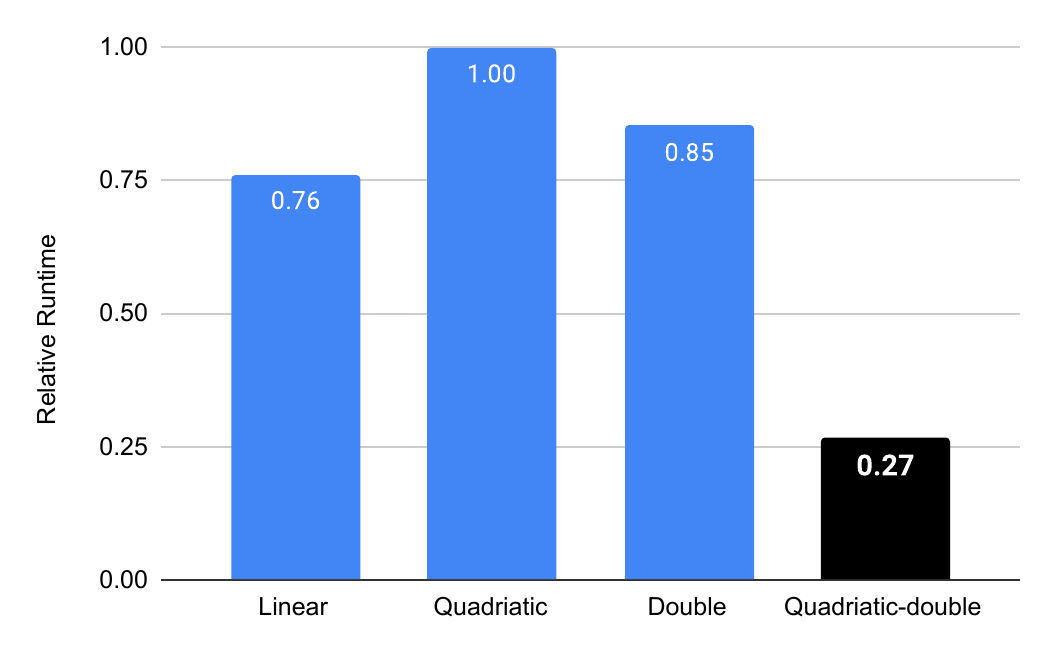}
  } \\[-2ex]
  \caption{Relative Runtime with using \textit{Linear} probing, \textit{Quadriatic} probing, \textit{Double} hashing, and a hybrid of Quadriatic probing and Double hashing (\textit{Quadriatic-double}) for collision resolution in the per-vertex hashtables\ignore{(which is used for storing the total associated weight with each label)}.}
  \label{fig:optresolution}
\end{figure}

We now examine three common techniques for resolving collisions in hashing: \textbf{linear probing}, \textbf{quadratic probing}, and \textbf{double hashing}. As previously mentioned, linear probing is cache-efficient but tends to experience high clustering, increasing the likelihood of repeated collisions. In contrast, double hashing significantly reduces clustering but lacks cache efficiency. Quadratic probing falls somewhere between the two in terms of clustering and cache efficiency. It is important to note that repeated collisions can lead to thread divergence, which can severely impact algorithm performance\ignore{--- whether using a thread-per-vertex or block-per-vertex approach}. In our implementation, we set the probe step for linear probing to $1$. For quadratic probing, we begin with an initial probe step of $1$ and double it with each subsequent collision. For double hashing, we use a secondary prime, $p_2 = nextPow2(p_1) - 1$, which is co-prime with $p_1 = |H|$. This secondary prime, $p_2$, is utilized with the modulo operator to calculate the probe step for each key. To further reduce collisions, we consider a hybrid collision resolution technique, which we refer to as \textbf{quadratic-double}. In this method, the probe step is the sum of the steps calculated by quadratic probing and double hashing. Figure \ref{fig:optresolution} compares the mean relative runtime of our GPU implementation of LPA using these four collision resolution strategies, with the per-vertex hashtables, on the large graphs presented in Table \ref{tab:dataset}.

As Figure \ref{fig:optresolution} shows, the \textbf{quadratic-double} approach delivers the best performance. Specifically, it is $2.8\times$, $3.7\times$, and $3.2\times$ times faster than LPA implementations that use linear probing, quadratic probing, and double hashing, respectively. This suggests that the hybrid approach effectively balances clustering and cache efficiency. Consequently, we adopt the quadratic-double strategy for our GPU-based LPA implementation. We also tested a coalesced chaining-based hashtable --- a collision resolution technique that combines aspects of separate chaining and open addressing --- utilizing another \textit{nexts} array $H_n$\ignore{for maintaining the chains needed}. However, it did not improve performance. We also experimented with shared memory-based hashtables for low-degree vertices, but saw little to no performance gain.

\subsection{Partitioning work between two kernels}

Processing each vertex in the graph with a thread-block per vertex may not be efficient, especially for low-degree vertices. If a vertex has a degree lower than $32$ --- the warp size on NVIDIA GPUs --- many threads in the warp would remain idle. Since graphs often contain a large number of such low-degree vertices, we opt to handle them using a thread-per-vertex approach\ignore{, a standard practice in the literature} \cite{wu2010efficient}. 

\begin{figure}[hbtp]
  \centering
  \subfigure{
    \label{fig:optswitchdegree--runtime}
    \includegraphics[width=0.98\linewidth]{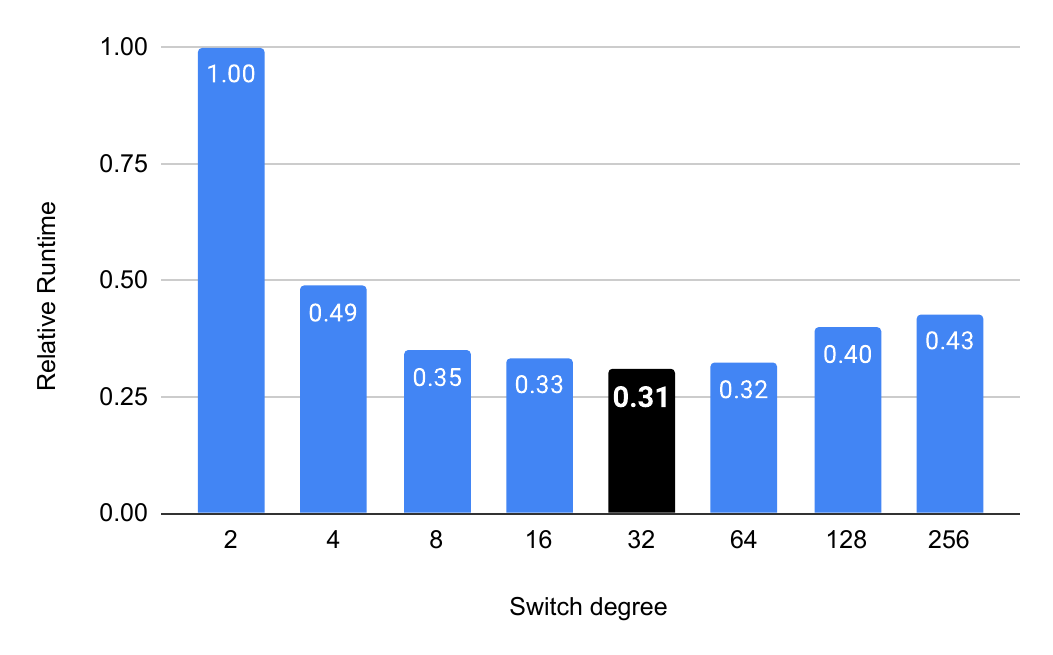}
  } \\[-2ex]
  \caption{Relative Runtime with various switch degrees, i.e., switching point by vertex degree between the thread-per-vertex kernel vs. block-per-vertex kernel, ranging from $2$ to $256$. Vertices with degree lower than the switch degree are processed by the thread-per-vertex kernel, while the remaining vertices are processed by the block-per-vertex kernel (the vertices are partitioned accordingly).}
  \label{fig:optswitchdegree}
\end{figure}

To implement this, we divide the vertices into two partitions: low-degree and high-degree. Low-degree vertices are processed using a thread-per-vertex kernel, while high-degree vertices are handled with a block-per-vertex kernel. We experiment with different partition thresholds, ranging from a degree of $2$ to $256$, using graphs from our dataset. As shown in Figure \ref{fig:optswitchdegree}, a \textbf{switch degree} of $32$ yields the best performance (highlighted in the figure). Note that in the thread-per-vertex kernel, only a single thread operates on the hashtable. This eliminates the need for atomic operations.

\subsection{Selecting datatype for hashtable values}

\begin{figure}[hbtp]
  \centering
  \subfigure{
    \label{fig:optdatatype--runtime}
    \includegraphics[width=0.98\linewidth]{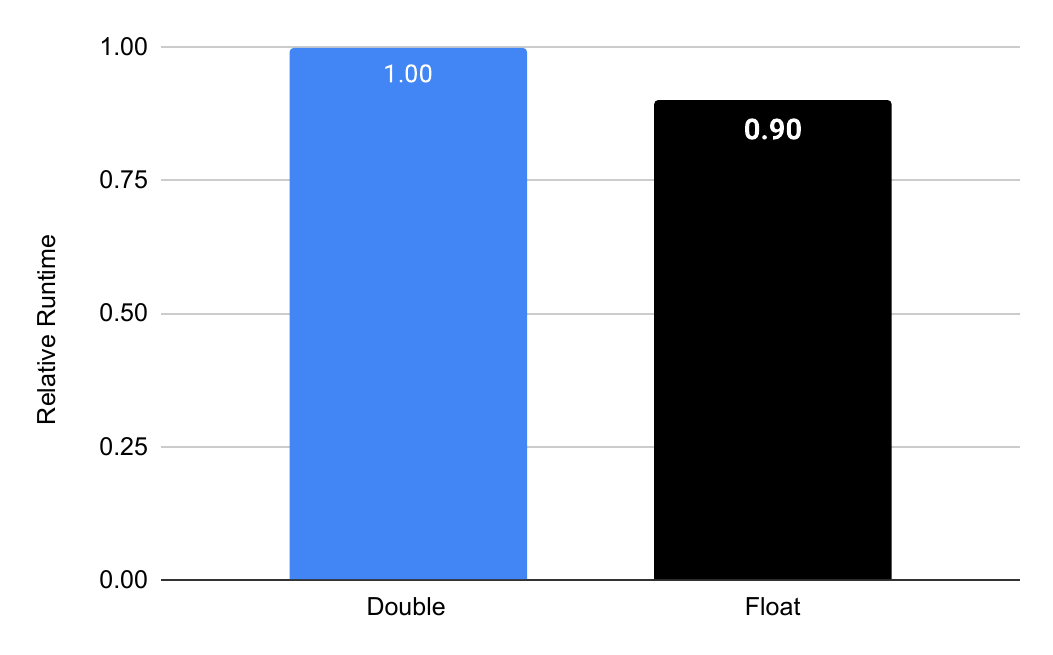}
  } \\[-2ex]
  \caption{Relative Runtime with using 32-bit floating point values (\textit{Float}) compared to 64-bit floating point values (\textit{Double}) for the aggregated weights (values) in the hashtable.}
  \label{fig:optdatatype}
\end{figure}

Finally, we experiment with using 32-bit floating-point numbers as hashtable values, i.e., for aggregated label weights, instead of 64-bit floating-point numbers. Figure \ref{fig:optdatatype} illustrates the relative runtime comparison between the two. Our findings indicate that using 32-bit floats does not compromise the quality of the communities obtained, while providing a moderate speedup. As a result, we adopt 32-bit floats as hashtable values in our GPU implementation of LPA.

\subsection{Our optimized LPA implementation}

We now explain our GPU implementation of LPA, which we refer to as \ours{}, in Algorithms \ref{alg:rak} and \ref{alg:hashtable}. In \ours{}, the symbol $\nu$ is a reference to ``video card". \ours{} has a time complexity of $O(KM)$, where $K$ is the number of iterations performed, and a space complexity of $O(M)$. In contrast, GVE-LPA has a space complexity of $O(TN + M)$, where $T$ represents the number of threads being used.

\subsubsection{Main step of \ours{}}

The main step of \ours{} is described in Algorithm \ref{alg:rak}. Here, the function \texttt{lpa()} takes the input graph $G$ and outputs the community membership (or label) for each vertex. In lines \ref{alg:rak--init-begin} to \ref{alg:rak--init-end}, we start by assigning a unique community label to each vertex --- effectively making each vertex its own community --- and mark all vertices as unprocessed. Next, multiple iterations of label propagation are carried out (lines \ref{alg:rak--iters-begin} to \ref{alg:rak--iters-end}). As discussed earlier, we use the \textit{Pick-Less (PL)} method every $\rho = 4$ iterations, where a vertex can only switch to a new community if the new community ID is smaller than its current one (lines \ref{alg:rak--pl-begin} to \ref{alg:rak--pl-end}) --- this helps mitigate community swaps. During each iteration, unprocessed vertices choose the label with the highest interconnecting weight via the \texttt{lpaMove()} function (line \ref{alg:rak--doiter}). In line \ref{alg:rak--converged}, we verify whether the fraction of changed vertices, $\Delta N/N$, is below the tolerance threshold $\tau$ and PL is disabled. If it is, the labels have converged, and the algorithm terminates, returning the identified communities in line \ref{alg:rak--return}.

Each iteration of \ours{} is performed in the \texttt{lpaMove()} function. Here, we first initialize the total count of changed vertices, $\Delta N$, the number of changed vertices per thread or thread-block $\Delta N_T$, and allocate space for the buffers $buf_k$ and $buf_v$ (each of size $2|E|$) for the per-vertex hash key arrays $H_k$ and hash value arrays $H_v$ (lines \ref{alg:rak--iterinit-begin}-\ref{alg:rak--iterinit-end}). Next, we perform the parallel update of labels for each vertex $i$ in the graph (lines \ref{alg:rak--iter-begin}-\ref{alg:rak--iter-end}). First, we mark vertex $i$ as processed. Then, if the degree of $i$ is less than \textit{SWITCH\_DEGREE} (set to 32), it is processed by an individual thread. Otherwise, a thread-block is used. For each vertex $i$, we calculate the size $|H| = p_1$ of the hashtable for $i$, and select a secondary prime number for quadratic-double hashing. The hashtable's memory, $H_k$ and $H_v$, is identified from the buffers $buf_k$ and $buf_v$ (lines \ref{alg:rak--hashtable-begin}-\ref{alg:rak--hashtable-end}). In line \ref{alg:rak--hashclear}, we first clear the hashtable in parallel, preparing for the identification of the most weighted label for $i$. Subsequently, we iterate over each neighbor $j$ of vertex $i$, accumulating its weighted label $(C[j], w)$ into the hashtable $H$ (lines \ref{alg:rak--scan-begin}-\ref{alg:rak--scan-end}). Afterward, we determine the most weighted label $c^*$ for vertex $i$ by performing a parallel max-reduce operation on the hashtable (line \ref{alg:rak--maxkey}). If the PL method is disabled or $c^*$ is smaller than $i$'s previous label, vertex $i$ adopts the new label $c^*$, the count of changed vertices per thread is updated, and the neighbors of vertex $i$ are marked as unprocessed in parallel (lines \ref{alg:rak--domove-begin}-\ref{alg:rak--domove-end}). Once all vertices assigned to the current thread/thread-block are processed, the global count of changed vertices $\Delta N$ is updated using atomic operations (line \ref{alg:rak--updatechanged}), and returned (line \ref{alg:rak--returnchanged}). Note that the community labels for vertices are updated in-place.

\begin{algorithm}[hbtp]
  \caption{\ours{}: Our GPU-based LPA.}
  \label{alg:rak}
  \begin{algorithmic}[1]
  \Require{$G(V, E)$: Input graph}
  \Require{$C$: Community membership of each vertex}
  \Ensure{$\Delta N$: Number of changed vertices, overall}
  \Ensure{$\Delta N_T$: Changed vertices per thread / thread-block}
  \Ensure{$c^*$: Most weighted label for vertex $i$}
  \Ensure{$H$: Per thread / thread-block hashtable\ignore{($H_k$: keys, $H_v$: values)}}
  \Ensure{$p_1$: Capacity of $H$, and also a prime}
  \Ensure{$p_2$: Secondary prime, such that $p_2 > p_1$}
  \Ensure{$l_i$: Number of iterations performed}
  \Ensure{$\tau$: Per iteration tolerance}

  \Statex

  \Function{lpa}{$G$}
    \State Vertex membership: $C \gets [0 .. |V|)$ \label{alg:rak--init-begin}
    \State Mark all vertices in $G$ as unprocessed \label{alg:rak--init-end}
    \ForAll{$l_i \in [0\ \dots\ \text{\small{MAX\_ITERATIONS}})$} \label{alg:rak--iters-begin}
      \State $\rhd$ Mitigate community swaps with \textbf{pick-less} mode
      \If{$l_i \bmod \rho = 0$} Enable \textbf{pick-less} mode \label{alg:rak--pl-begin}
      \Else\ Disable \textbf{pick-less} mode
      \EndIf \label{alg:rak--pl-end}
      \State $\Delta N \gets lpaMove(G, C)$ \label{alg:rak--doiter}
      \If{\textbf{not pick-less and} $\Delta N / N < \tau$} \textbf{break} \label{alg:rak--converged}
      \EndIf
    \EndFor \label{alg:rak--iters-end}
    \Return{$C$} \label{alg:rak--return}
  \EndFunction

  \Statex

  \Function{lpaMove}{$G, C$}
    \State $\Delta N \gets 0$ \textbf{;} $\Delta N_T \gets \{\}$ \label{alg:rak--iterinit-begin}
    \State $buf_k \gets \{\}$ \textbf{;} $buf_v \gets \{\}$
    \State $\Delta N_T[t] \gets 0$ \textbf{on each thread / thread-block} \label{alg:rak--iterinit-end}
    \ForAll{\textbf{unprocessed} $i \in V$ \textbf{in parallel}} \label{alg:rak--iter-begin}
      \State $\rhd$ Mark vertex $i$ as processed
      \State $\rhd$ If degree of $i <$ \small{SWITCH\_DEGREE},
      \State $\rhd$ process using a thread, else use a thread-block
      \State $p_1 \gets nextPow2(G.degree(i)) - 1$ \label{alg:rak--hashtable-begin}
      \State $p_2 \gets nextPow2(p_1) - 1$
      \State $\theta_H \gets 2 * G.offset(i)$
      \State $H_k \gets buf_k[\theta_H : \theta_H + p_1]$ \Comment{$H$ is \textbf{shared}, if using}
      \State $H_v \gets buf_v[\theta_H : \theta_H + p_1]$ \Comment{a thread-block} \label{alg:rak--hashtable-end}
      \State $\rhd$ Identify most weighted label for vertex $i$
      \State $hashtableClear(H)$ \textbf{in parallel} \label{alg:rak--hashclear}
      \ForAll{$(j, w) \in G.neighbors(i)$ \textbf{in parallel}} \label{alg:rak--scan-begin}
        \If{$j = i$} \textbf{continue}
        \EndIf
        \State $hashtableAccumulate(H, p_1, p_2, C[j], w)$ \label{alg:rak--accumulate}
      \EndFor \label{alg:rak--scan-end}
      \State $c^* \gets hashtableMaxKey(H)$ \textbf{in parallel} \label{alg:rak--maxkey}
      \State $\rhd$ Change label of vertex $i$ to most weighted label $c^*$
      \If{$c^* \neq C[i]$ \textbf{and} $($\textbf{not pick-less or} $c^* \leq C[i])$} \label{alg:rak--domove-begin}
        \State $C[i] \gets c^*$
        \State $\Delta N_T[t] \gets \Delta N_T[t] + 1$
        \ForAll{$j \in G.neighbors(i)$ \textbf{in parallel}}
          \State Mark $j$ as unprocessed
        \EndFor
      \EndIf \label{alg:rak--domove-end}
    \EndFor \label{alg:rak--iter-end}
    \State $atomicAdd(\Delta N, \Delta N_T[t])$ \textbf{on each thread / thread-block} \label{alg:rak--updatechanged}
    \Return{$\Delta N$} \label{alg:rak--returnchanged}
  \EndFunction
  \end{algorithmic}
  \end{algorithm}

\subsubsection{Hashtable for \ours{}}

We now turn our attention to Algorithm \ref{alg:hashtable}, which is designed to accumulate the associated weights of labels for a vertex within its hashtable, while efficiently handling potential collisions through a hybrid quadratic-double probing strategy. Given a key $k$ and its corresponding value $v$, the algorithm aims to locate an appropriate slot in the hashtable $H$ (with $p_1$ slots) where the key can either be inserted or its value updated. The process begins by calculating an initial slot index $s$ using the primary hash function: $s = i \bmod p_1$ (line \ref{alg:hashtable--slot}), where $i$ starts as the key $k$ and is adjusted incrementally using a step size $\delta i$. If the computed slot $s$ is empty (denoted by $\phi$) or already contains the key $k$, the algorithm proceeds to update or insert the value in $H_v[s]$.

The approach differs depending on whether the hashtable is accessed by multiple threads (shared scenario, lines \ref{alg:hashtable--shared-begin}-\ref{alg:hashtable--shared-end}) or a single thread (non-shared scenario, lines \ref{alg:hashtable--unshared-begin}-\ref{alg:hashtable--unshared-end}). In the non-shared case, the algorithm directly modifies $H_k$ and $H_v$. In the shared case, atomic operations ensure thread safety: an \texttt{atomicCAS()} operation attempts to atomically set $H_k[s]$ to $k$ if the slot is empty. If this operation succeeds, or if the slot already contains $k$, the \texttt{atomicAdd()} operation updates $H_v[s]$ by adding the new value $v$. If the slot is occupied by a different key, hybrid quadratic-double probing is triggered as follows: $\delta i$ is doubled and adjusted using $k \bmod p_2$ (line \ref{alg:hashtable--update-end}), where $p_2$ is a secondary prime larger than $p_1$. This process repeats iteratively (lines \ref{alg:hashtable--loop-begin}-\ref{alg:hashtable--loop-end}) for up to \textit{MAX\_RETRIES} attempts. If a suitable slot is not located within the maximum retries, the function returns a \textit{failed} status --- though this scenario is avoided by ensuring the hashtable has sufficient capacity for all entries.

\begin{algorithm}[hbtp]
\caption{Accumulating associated weights of labels.}
\label{alg:hashtable}
\begin{algorithmic}[1]
\Require{$H$: Hashtable with $p_1$ slots ($H_k$: keys, $H_v$: values)}
\Require{$p_1$: Capacity of $H$, and also a prime}
\Require{$p_2$: Secondary prime, such that $p_2 > p_1$}
\Require{$k, v$: Key, associated value to accumulate}
\Ensure{$s$: Slot index}

\Statex

\Function{hashtableAccumulate}{$H, p_1, p_2, k, v$}
  \State $i \gets k$ \textbf{;} $\delta i \gets 1$ \label{alg:hashtable--init}
  \ForAll{$t \in [0\ \dots\ \text{\small{MAX\_RETRIES}})$} \label{alg:hashtable--loop-begin}
    \State $s \gets i \bmod p_1$ \Comment{$1^{st}$ hash function} \label{alg:hashtable--slot}
    \If{\textbf{is not shared}}
      \If{$H_k[s] = k$ \textbf{or} $H_k[s] = \phi$} \label{alg:hashtable--shared-begin}
        \If{$H_k[s] = \phi$} $H_k[s] \gets k$
        \EndIf
        \State $H_v[s] \gets v$
        \Return{$done$}
      \EndIf \label{alg:hashtable--shared-end}
    \Else
      \If{$H_k[s] = k$ \textbf{or} $H_k[s] = \phi$} \label{alg:hashtable--unshared-begin}
        \State $k_{old} \gets atomicCAS(H_k[s], \phi, k)$
        \If{$k_{old} = \phi$ \textbf{or} $k_{old} = k$}
          \State $atomicAdd(H_v[s], v)$
          \Return{$done$}
        \EndIf
      \EndIf \label{alg:hashtable--unshared-end}
    \EndIf
    \State $i \gets i + \delta i$ \label{alg:hashtable--update-begin}
    \State $\delta i \gets 2 * \delta i + (k \bmod p_2)$ \Comment{$2^{nd}$ hash function}\label{alg:hashtable--update-end}
  \EndFor \label{alg:hashtable--loop-end}
  \Return{$failed$}
\EndFunction
\end{algorithmic}
\end{algorithm}

\section{Evaluation}
\label{sec:evaluation}
\subsection{Experimental Setup}
\label{sec:setup}

\subsubsection{System used}
\label{sec:system}

We utilize a server featuring an NVIDIA A100 GPU, which has 108 Streaming Multiprocessors (SMs), each with 64 CUDA cores. The GPU has $80$ GB of global memory, a bandwidth of $1935$ GB/s, and $164$ KB of shared memory per SM. The server is powered by an AMD EPYC-7742 processor with 64 cores running at $2.25$ GHz, has $512$ GB of DDR4 RAM, and runs Ubuntu 20.04. For evaluating CPU-only implementations of LPA, we use a separate server which has two Intel Xeon Gold 6226R processors, with each processor having 16 cores clocked at $2.90$ GHz. Each core of the CPU is equipped with a $1$ MB L1 cache, a $16$ MB L2 cache, and a $22$ MB shared L3 cache. The system is configured with $512$ GB of RAM and runs on CentOS Stream 8.

\subsubsection{Configuration}
\label{sec:configuration}

We utilize 32-bit integers for vertex identifiers and 32-bit floating-point numbers for edge weights and hash table values. For compilation, we employ GCC 9.4, OpenMP 5.0, and CUDA 11.4 on the GPU system. On the CPU-only system, we rely on GCC 8.5 and OpenMP 4.5.

\subsubsection{Dataset}
\label{sec:dataset}

The graphs used in our experiments are listed in Table \ref{tab:dataset}, and they are sourced from the SuiteSparse Matrix Collection \cite{suite19}. The graphs vary in size, with the number of vertices ranging from $3.07$ million to $214$ million, and the number of edges ranging from $25.4$ million to $3.80$ billion. We ensure that the edges are undirected and weighted, with a default weight of $1$.

\begin{table}[hbtp]
  \centering
  \caption{List of $13$ graphs obtained SuiteSparse Matrix Collection \cite{suite19} (directed graphs are marked with $*$). Here, $|V|$ is the number of vertices, $|E|$ is the number of edges (after adding reverse edges), and $D_{avg}$ is the average degree, and $|\Gamma|$ is the number of communities obtained with \textit{\ours{}}.}
  \label{tab:dataset}
  \begin{tabular}{|c||c|c|c|c|}
    \toprule
    \textbf{Graph} &
    \textbf{\textbf{$|V|$}} &
    \textbf{\textbf{$|E|$}} &
    \textbf{\textbf{$D_{avg}$}} &
    \textbf{\textbf{$|\Gamma|$}} \\
    \midrule
    \multicolumn{5}{|c|}{\textbf{Web Graphs (LAW)}} \\ \hline
    indochina-2004$^*$ & 7.41M & 341M & 41.0 & 215K \\ \hline
    uk-2002$^*$ & 18.5M & 567M & 16.1 & 541K \\ \hline
    arabic-2005$^*$ & 22.7M & 1.21B & 28.2 & 364K \\ \hline
    uk-2005$^*$ & 39.5M & 1.73B & 23.7 & 1.14M \\ \hline
    webbase-2001$^*$ & 118M & 1.89B & 8.6 & 8.51M \\ \hline
    it-2004$^*$ & 41.3M & 2.19B & 27.9 & 901K \\ \hline 
    sk-2005$^*$ & 50.6M & 3.80B & 38.5 & ? \\ \hline
    \multicolumn{5}{|c|}{\textbf{Social Networks (SNAP)}} \\ \hline
    com-LiveJournal & 4.00M & 69.4M & 17.4 & 145K \\ \hline
    com-Orkut & 3.07M & 234M & 76.2 & 2.21K \\ \hline
    \multicolumn{5}{|c|}{\textbf{Road Networks (DIMACS10)}} \\ \hline
    asia\_osm & 12.0M & 25.4M & 2.1 & 2.01M \\ \hline
    europe\_osm & 50.9M & 108M & 2.1 & 7.51M \\ \hline
    \multicolumn{5}{|c|}{\textbf{Protein k-mer Graphs (GenBank)}} \\ \hline
    kmer\_A2a & 171M & 361M & 2.1 & 28.8M \\ \hline
    kmer\_V1r & 214M & 465M & 2.2 & 34.7M \\ \hline
  \bottomrule
  \end{tabular}
\end{table}

\subsection{Performance Comparison}
\label{sec:performance-comparison}  

We now evaluate the performance of \ours{}, comparing it with existing implementations. Besides Gunrock LPA \cite{wang2016gunrock}, no other functional GPU-based LPA implementations are publicly available. Therefore, as discussed earlier, we compare \ours{} with state-of-the-art sequential implementations like FLPA \cite{traag2023large} and multicore implementations such as NetworKit LPA \cite{staudt2016networkit}. Additionally, we contrast \ours{} with cuGraph Louvain \cite{kang2023cugraph}, a state-of-the-art GPU-based implementation of the Louvain algorithm, to highlight the trade-offs between LPA and the Louvain algorithm.

\begin{figure*}[hbtp]
  \centering
  \subfigure[Runtime in seconds (logarithmic scale) with \textit{FLPA}, \textit{NetworKit LPA}, \textit{Gunrock LPA}, \textit{cuGraph Louvain}, and \textit{\ours{}}]{
    \label{fig:compare--runtime}
    \includegraphics[width=0.98\linewidth]{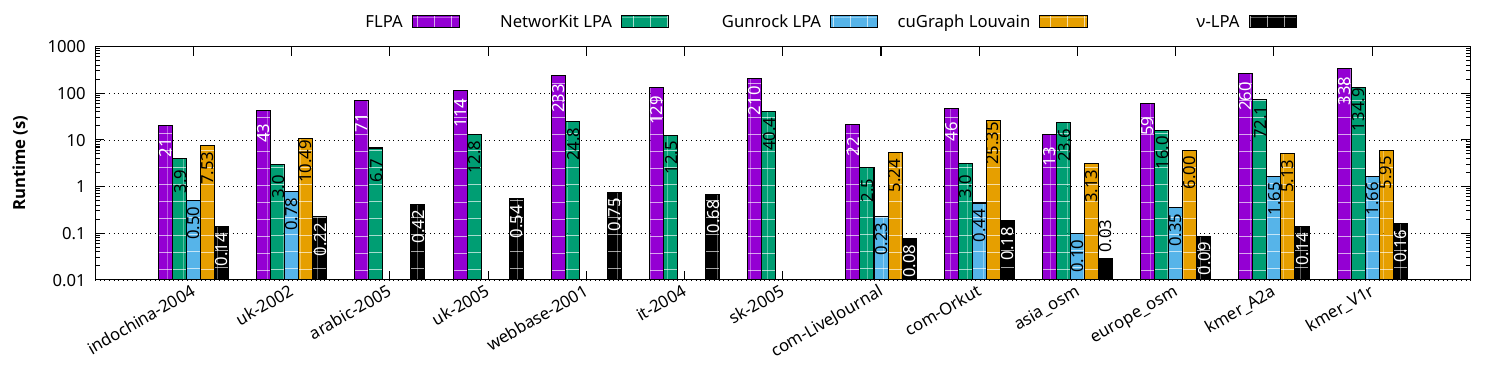}
  }
  \subfigure[Speedup of \textit{\ours{}} (logarithmic scale) with respect to \textit{FLPA}, \textit{NetworKit LPA}, \textit{cuGraph Louvain}, and \textit{Gunrock LPA}.]{
    \label{fig:compare--speedup}
    \includegraphics[width=0.98\linewidth]{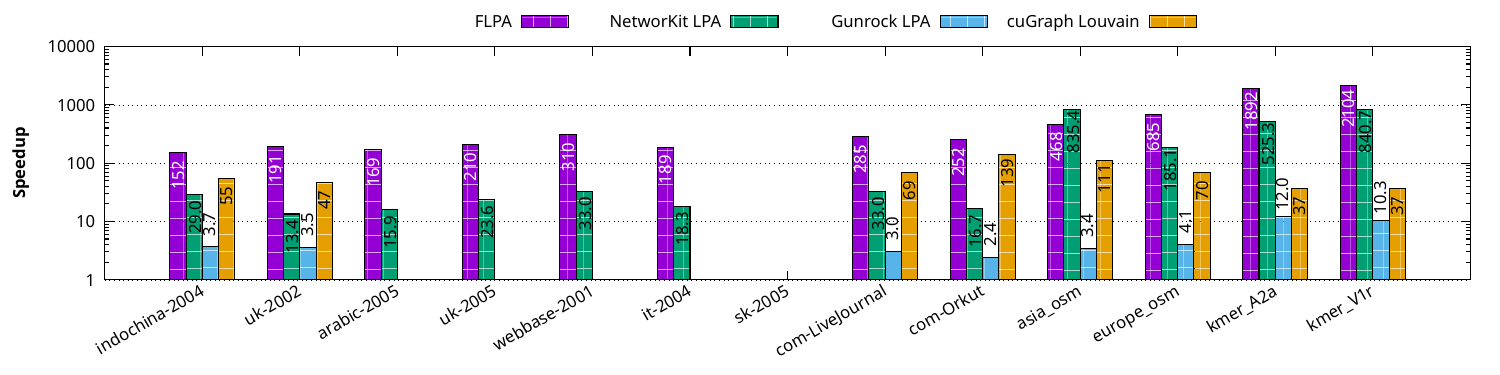}
  }
  \subfigure[Modularity of communities obtained with \textit{FLPA}, \textit{NetworKit LPA}, \textit{Gunrock LPA}, \textit{cuGraph Louvain}, and \textit{\ours{}}.]{
    \label{fig:compare--modularity}
    \includegraphics[width=0.98\linewidth]{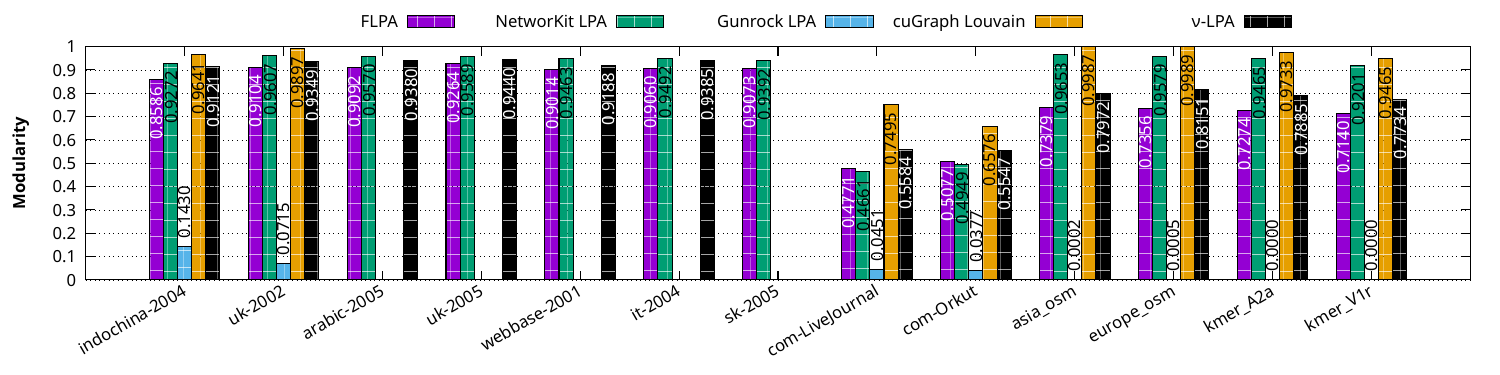}
  } \\[-2ex]
  \caption{Runtime in seconds (log-scale), speedup (log-scale), and modularity of obtained communities with \textit{FLPA}, \textit{NetworKit LPA}, \textit{Gunrock LPA}, \textit{cuGraph Louvain}, and \textit{\ours{}} for each graph in the dataset.}
  \label{fig:compare}
\end{figure*}

For FLPA, we check out the appropriate branch containing the modified \texttt{igraph\_community\_label\_propagation()} function, update the label propagation example in C to load the input graph from a file, and measure the runtime of \texttt{igraph\_community\_label\_pr} \texttt{opagation()} using the \texttt{IGRAPH\_LPA\_FAST} variant and \texttt{gettimeof} \texttt{day()}. For NetworKit, we utilize a Python script to execute \texttt{PLP} (Parallel Label Propagation) and measure the total runtime with \texttt{getTiming()}. For Gunrock LPA, we download version 0.5.1, run \texttt{LpProblem}, and measure the only iteration time using \texttt{cpu\_timer}. To test cuGraph Louvain, we write a Python script that configures the Rapids Memory Manager (RMM) to use a pool allocator for faster memory allocations, then run \texttt{cugraph.louvain()} on the graph. We repeat the runtime measurement five times per graph to obtain an average, and record the modularity of communities identified by each implementation. When using cuGraph, we disregard the runtime of the first run to ensure subsequent measurements accurately reflect RMM's pool usage without the initial CUDA memory allocation overhead.

Figure \ref{fig:compare--runtime} presents the runtimes of FLPA, NetworKit LPA, Gunrock LPA, cuGraph Louvain, and \ours{} for each graph in the dataset, while Figure \ref{fig:compare--speedup} illustrates the speedup of \ours{} relative to the other implementations. Note that both Gunrock LPA and cuGraph Louvain fail to run on the \textit{arabic-2005}, \textit{uk-2005}, \textit{webbase-2001}, \textit{it-2004}, and \textit{sk-2005} graphs, while \ours{} fails to run on the \textit{sk-2005} graph --- due to out-of-memory issues --- so these results are omitted.

\ours{} demonstrates an average speedup of $364\times$, $62\times$, $2.6\times$, and $37\times$ compared to FLPA, NetworKit LPA, Gunrock LPA, and cuGraph Louvain, respectively. On the \textit{it-2004} graph, \ours{} identifies communities in just $1.6$ seconds, achieving a processing rate of $3.0$ billion edges per second. Figure \ref{fig:compare--modularity} shows the modularity of the communities identified by each method. On average, \ours{} achieves $4.7\%$ higher modularity than FLPA, particularly on road networks and protein k-mer graphs. However, it produces $6.1\%$ lower modularity than NetworKit LPA and $9.6\%$ lower modularity than cuGraph Louvain on similar graphs. Notably, the modularity achieved by Gunrock LPA is very low. Addressing the modularity gap relative to NetworKit LPA remains a focus for future work. Despite this, the results highlight \ours{} as a strong contender for web graphs and social networks, offering significant performance improvements over existing state-of-the-art implementations.

\section{Conclusion}
\label{sec:conclusion}
In conclusion, in this technical report, we introduced an efficient implementation of the Label Propagation Algorithm (LPA) for community detection. It used an asynchronous implementation of LPA, which employed a Pick-Less (PL) approach every $4$ iterations for handling community swaps --- a scenario more likely to occur on SIMT hardware, such as GPUs. Additionally, it utilized a novel open-addressing-based per-vertex hashtable with hybrid quadratic-double probing for collision resolution. On a server with an NVIDIA A100 GPU, our GPU implementation of LPA, referred to as \ours{}, outperformed FLPA, NetworKit LPA, Gunrock LPA, and cuGraph Louvain by $364\times$, $62\times$, $2.6\times$, and $37\times$, respectively, running on a server equipped with dual 16-core Intel Xeon Gold 6226R processors --- achieving a processing rate of $3.0B$ edges/s on a $2.2B$ edge graph --- while identifying communities with $4.7\%$ higher modularity than FLPA, but $6.1\%$ and $9.6\%$ lower modularity than NetworKit LPA and cuGraph Louvain. This highlights the applicability of \ours{} for performance-critical applications, such as partitioning of large graphs. We plan to look into this in the future.

\begin{acks}
I would like to thank Prof. Kishore Kothapalli, Prof. Dip Sankar Banerjee, and Souvik Karfa for their support.
\end{acks}

\bibliographystyle{ACM-Reference-Format}
\bibliography{main}

\end{document}